\documentclass[12pt]{article}

\usepackage[latin2]{inputenc}
\usepackage{amsmath}
\usepackage{amssymb}
\usepackage{amsfonts}
\usepackage{stmaryrd}
\usepackage{theorem}
\usepackage{graphicx}
\usepackage{epstopdf}
\usepackage[parfill]{parskip}    
\usepackage{hyperref}
\hypersetup{pdftex,colorlinks=true,allcolors=blue}
\usepackage{hypcap}
\usepackage{algorithm}
\usepackage{algpseudocode} 
\usepackage{authblk}
\usepackage{tabto}
\usepackage{wrapfig} 

\newcommand{\libeq}{\mathrel{\mathop:}=}
\newcommand*{\me}{\mathrm{e}}
\newcommand*{\vi}{\mathrm{i}}
\newcommand*{\mmod}{\mathrm{mod}}
\DeclareMathOperator*{\mamin}{\mathrm{argmin}}

\newcommand*{\Conv}{\mathrm{Conv}}

\newcommand*{\Ext}{\mathrm{Ext}}

\newcommand*{\mRe}{\mathrm{Re}}
\newcommand*{\mIm}{\mathrm{Im}}

\newcommand*{\mH}{\mathrm{H}}
\newcommand*{\mM}{\mathrm{M}}

\newcommand*{\cf}{\ensuremath{\varphi}}
\newcommand*{\vr}{\ensuremath{\varrho}}
\newcommand*{\ve}{\ensuremath{\varepsilon}}
\newcommand*{\vt}{\ensuremath{\vartheta}}

\newcommand*{\vl}{\ensuremath{\lambda}}
\newcommand*{\va}{\ensuremath{\alpha}}

\newcommand*{\N}{\ensuremath{\mathbb{N}}}
\newcommand*{\Z}{\ensuremath{\mathbb{Z}}}

\newcommand*{\R}{\ensuremath{\mathbb{R}}}
\newcommand*{\C}{\ensuremath{\mathbb{C}}}

\newcommand*{\lrar}{\ensuremath{\Leftrightarrow}}

\newcommand*{\co}{\ensuremath{\circ}}
\newcommand*{\bu}{\ensuremath{\bullet}}
\newcommand*{\tti}{\ensuremath{\rightarrow\infty}}

\begin{document}

\title{Apollonian Circumcircles of IFS Fractals}
\author{J\'ozsef Vass}
\affil{Faculty of Mathematics\\ University of Waterloo\\ \vspace{0.35cm} jvass@uwaterloo.ca}
\date{}
\maketitle

\begin{abstract}
\noindent Euclidean triangles and IFS fractals seem to be disparate geometrical concepts, unless we consider the Sierpi\'{n}ski gasket, which is a self-similar collection of triangles. The ``circumcircle'' hints at a direct link, as it can be derived for three-map IFS fractals in general, defined in an Apollonian manner. Following this path, one may discover a broader relationship between polygons and IFS fractals.\footnote{Previously titled ``Explicit Bounding Circles for IFS Fractals''. Some of these results appeared in the author's doctoral dissertation \cite{ph00004}.}
\end{abstract}

\vspace{0.5cm}
\tableofcontents
\newpage

\section{Introduction}

The Sierpi\'{n}ski gasket evolves as a limit set by iteratively shrinking an equilateral triangle towards its three vertices, as illustrated below.
\vspace{0.375cm}
\begin{figure}[H]
\centering
\includegraphics[width=470pt]{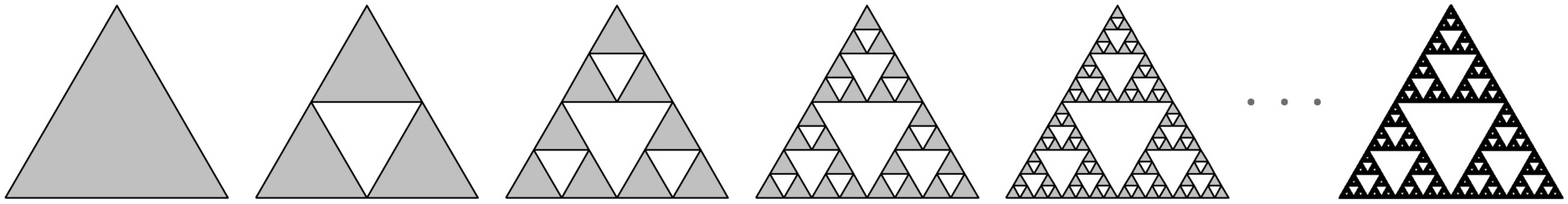}
\vspace{-0.25cm}
\caption{This converges to an attractor with Hausdorff dimension $\log_2 3$ (hence ``fractal'').}
\end{figure}
\vspace{-0.225cm}
This iteration can be viewed as the collective action of three contractive affine transformations $T_k$ with contraction factors $\vl_k\in (0,1)$ and fixed points $p_k\in\C$ at the vertices. Adding rotations $\vt_k\in (-\pi,\pi]$ to the actions of $T_k$ for a little more generality, their trajectories will be logarithmic spirals, and will take the form
\[ T_k(z)=p_k+\cf_k(z-p_k)\ \ (z\in\C,\ k=1,2,3) \]
where $\cf_k=\vl_k\me^{\vt_k\vi}$ and their collective action can be represented by the map
\[ \mH(S)\libeq T_1(S)\cup T_2(S)\cup T_3(S) \]
which has a unique attractor $F=\mH(F)$ over compact sets as shown by Hutchinson \cite{ba00007}, commonly called an ``IFS fractal'' where IFS stands for ``iterated function system''. In this sense, three-map IFS fractals are generalized triangular fractals, or ``trifractals''. Their definition can of course be generalized to any dimension $d\ge 1$ with one-or-more contractions, not far removed from Nature considering that the Romanesco broccoli is a 3D IFS fractal, due to botanical L-systems being close relatives of such fractals \cite{bb00005}.

\section{Apollonian Circumcircles}

\subsection{Trifractals}

Let us consider the vague problem of generalizing the Euclidean circumcircle to trifractals. Observing that due to the scaling action of $T_k$ the circumcircle of the Sierpi\'{n}ski gasket $F$ is tangential to the circumcircles of each subfractal $T_k(F)$, we might attempt an analogous Apollonian definition in general. The conditions for inner tangentiality of a circle $C=(c,r)\in\C\times\R_+$ with its iterates $T_k(C)$ are
\[ |T_k(c)-c|+\vl_k r = r\ \ (k=1,2,3). \]
\begin{figure}[H]
\centering
\includegraphics[width=420pt]{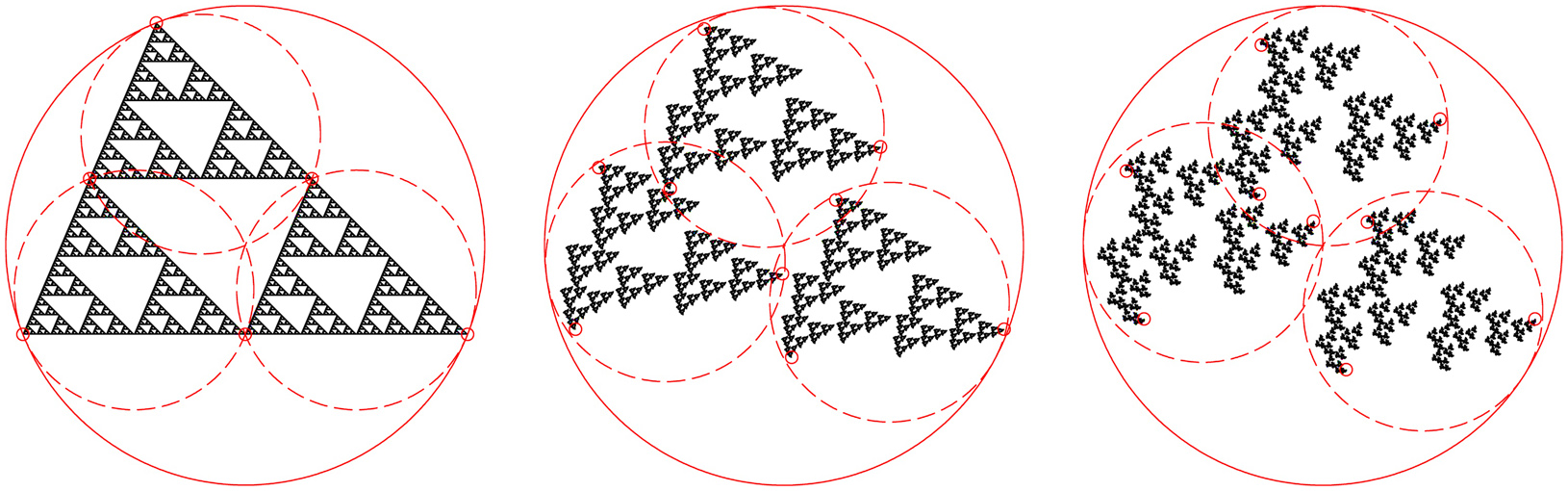}
\caption{The circumcircle varying under perturbation of the IFS rotations.}
\end{figure}
\vspace{-0.4cm}
These three equations in three real unknowns $r, \mRe(c), \mIm(c)$ can be reduced to
\[ |p_k-c|^2-\va_k^2 r^2 =0\ \ \mathrm{with}\ \ \va_k\libeq \frac{1-|\cf_k|}{|1-\cf_k|}\ \ \ (k=1,2,3). \]
Denoting $p_{k1}\libeq \mRe(p_k),\ p_{k2}\libeq \mIm(p_k),\ x\libeq \mRe(c),\ y\libeq \mIm(c)$ and expanding these conditions, then multiplying each by the factors $p_{31}-p_{21},\ p_{11}-p_{31},\ p_{21}-p_{11}$ respectively, and lastly summing the three equations, several terms drop out in the resulting equation, which we denote as $E_1=0$. We may do similarly with the factors $p_{32}-p_{22},\ p_{12}-p_{32},\ p_{22}-p_{12}$, resulting in an analogous equation $E_2=0$. Taking $E_1+E_2\vi=0$ and collecting terms, we get an equation of the following form with these constants
\[ A r^2+B-(C\vi)c = 0 \]
\[ A\libeq (\va_3^2-\va_2^2)p_1 + (\va_1^2-\va_3^2)p_2 + (\va_2^2-\va_1^2)p_3 \]
\[ B\libeq (|p_2|^2-|p_3|^2)p_1 + (|p_3|^2-|p_1|^2)p_2 + (|p_1|^2-|p_2|^2)p_3 \]
\[ C\libeq 2(p_2-p_1)\times(p_2-p_3) \]
with the complex cross product $z_1\times z_2\libeq \mRe(z_1)\mIm(z_2)-\mIm(z_1)\mRe(z_2)$ for $z_{1,2}\in\C$.

The above equation is solvable for the center $c$ in terms of the radius $r$ iff $C\neq 0$, meaning iff the fixed points $p_{1,2,3}$ are non-collinear. In that case, the center takes the form $c=c_0+a r^2$ with $c_0=B/C\vi,\ a=A/C\vi$. If the fractal were Sierpi\'{n}ski, then $\vt_k=0$ and thus $\va_k=1$ held for each $k$, implying $A=0$. So $c_0$ is the center of the circumcircle of $p_{1,2,3}$ and let its radius be $r_0\libeq |p_1-c_0|$.

To find the fractal circumcircle radius $r$, let us expand one of the tangentiality conditions, say the first one $|p_1-c|^2-\va_1^2 r^2 =0$, resulting after some algebraic manipulation in the equation
\[ |a|^2 r^4 + 2D r^2 + r_0^2 =0 \]
\[ D\libeq a\bu c_0 + \frac{1}{C}(\va_1^2\ p_2\times p_3 +\va_2^2\ p_3\times p_1 + \va_3^2\ p_1\times p_2) = a\bu c_0 -\left(a\bu p_1 + \frac{\va_1^2}{2}\right) \]
where $\bu$ is the dot product for complex vectors.

If $a=0$ ($\lrar A=0$), then $\va_{1,2,3}$ are equal since $p_{1,2,3}$ are non-collinear, so $r=r_0/\va_1$ and $c=c_0$. On the other hand, if $a\neq 0$ then we can solve the above equation as a quadratic for $r^2$. Taking the square root we get
\[ r=\frac{1}{|a|}\sqrt{-D\pm\sqrt{D^2- (|a| r_0)^2}} \]
implying two solution circles if $p_{1,2,3}$ are non-collinear and $D\le -|a|r_0$, with $c= c_0+ a r^2$. The smaller one may be preferable to be defined as ``the circumcircle'' of a trifractal.

Note that if $\vt_k=0$ for some $k\in\{1,2,3\}$, then $\va_k=1$, so by the corresponding circumcircle condition, we get that $|p_k-c|=r$. This implies that the corresponding fixed point $p_k$ lies on the circumcircle. In fact for Sierpi\'{n}ski trifractals with $\vt_k=0\ \forall k$, all three fixed points lie on the circumcircle, as expected.

\subsection{Bifractals}

\begin{wrapfigure}{r}{0.5\textwidth}
\begin{center}
\vspace{-0.4cm}
\includegraphics[width=0.46\textwidth]{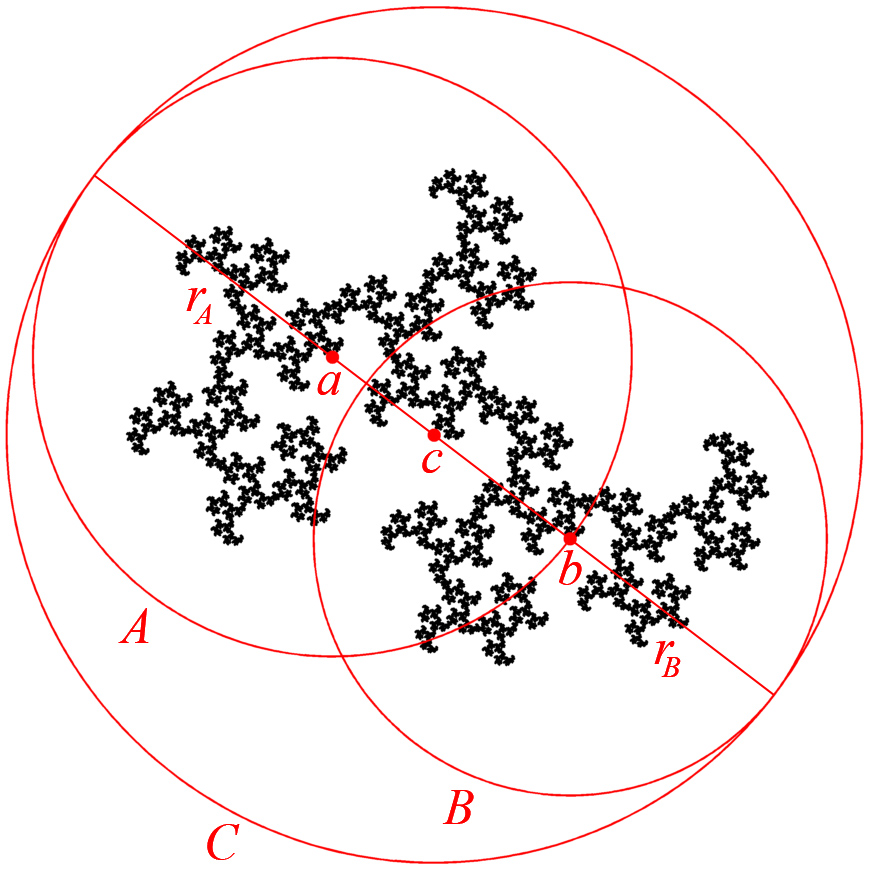}
\end{center}
\vspace{0.35cm}
\caption{The circumcircle of a bifractal.}
\end{wrapfigure}
In the case of two IFS contractions, if the rotations are both zero then the attractor is a Cantor set along the segment connecting the two fixed points. The one-dimensional ``bounding ball'' is the segment itself, while the two-dimensional one is a circular disk centered at the midpoint. We might wonder if the latter can be generalized, possibly again in an Apollonian manner.

In our attempt, we take a completely different approach than for trifractals, based on the following intuitive figure depicting the sought circumcircle $C=(c,r)$ and its iterates $A=(a,r_A)$ and $B=(b,r_B)$ according to the contractions $T_1$ and $T_2$.

Based on the figure, we see that the following equations must hold:\\
\[ c = \frac{1}{2}((a - r_A u) + (b + r_B u)),\ \ r = \frac{1}{2}(r_A + r_B + |b-a|)\ \ \mathrm{with}\ \ u\libeq \frac{b-a}{|b-a|}. \]
Since $T_1$ maps $C$ to $A$ and $T_2$ maps $C$ to $B$, we have that $A=(a,r_A)=(T_1(c), \vl_1 r)$ and $B=(b,r_B)=(T_2(c), \vl_2 r)$. So defining $\mM=(m_1,m_2):\C\times\R_+\shortrightarrow\C\times\R_+$ as
\[ m_1(c,r)\libeq \frac{T_1(c)+T_2(c)}{2} + r\frac{\vl_2-\vl_1}{2}\frac{T_2(c)-T_1(c)}{|T_2(c)-T_1(c)|},\ \ m_2(c,r)\libeq \frac{\vl_1+\vl_2}{2}r + \frac{|T_2(c)-T_1(c)|}{2} \]
for $T_1(c)\neq T_2(c)$ the sought circumcircle will be its fixed point $(c,r)=\mM(c,r)$.

For the second component $m_2(c,r)=r$ we get that $r=|T_2(c)-T_1(c)|/2(1-\vl)$ denoting $\lambda\libeq (\lambda_1+\lambda_2)/2$. Plugging this into the first fixed point equation $m_1(c,r)=c$ we get with $\nu\libeq (\lambda_2-\lambda_1)/2(1-\lambda)$ the formula for the center
\[ c=\ \frac{(1-\nu)(1-\cf_1)p_1 + (1+\nu)(1-\cf_2)p_2}{(1-\nu)(1-\cf_1) +(1+\nu)(1-\cf_2)} \]
and remarkably it is a ``complex combination'' of the fixed points $p_{1,2}$. Plugging this formula for $c$ into the expression $r=|T_2(c)-T_1(c)|/2(1-\vl)$ we get that
\[ r=\ \frac{1}{1-\lambda}\ \frac{|1-\varphi_1|\ |1-\varphi_2|}{|(1-\nu)(1-\varphi_1) +(1+\nu)(1-\varphi_2)|}\ |p_2-p_1|. \]
This formula implies that $r\neq 0$ (since $p_1=p_2$ would degenerate the fractal to a point) so by its earlier relationship to $|T_2(c)-T_1(c)|$ we see that $T_1(c)\neq T_2(c)$. Therefore this derivation shows that the fixed point of $\mM$ exists and it is unique. Note that by further investigation, we find that $\mM$ is not a contractive map in general.

\subsection{Polyfractals}

In the case of more than two IFS contractions -- ``polyfractals'' -- it may be tempting to consider when the convex hull of the IFS fixed points is a cyclic polygon. Perhaps its circumcircle could be ``blown up'' as for trifractals.

For now assume instead that the IFS rotations are all equal -- ``equiangular'' -- and of the form $\vt\libeq \vt_k=2\pi N/M,\ M\in\N,\ N\in [0,M)\cap\Z$. Such a fractal $F$ generated by the contractions $T_1,\ldots,T_n$ can also be generated as a Sierpi\'{n}ski fractal. Meaning with a new IFS, we can generate the same fractal but with zero rotations.

To see this, first of all let us observe that the identity $F=\mH(F)$ inductively implies that
\[ F=\mH(F)=\ldots=\mH^L(F)\ \ (L\in\N). \]
Notice that due to the definition of $\mH$ ``addresses'' $a$ of length $L$ (denoted $|a|=L$) are generated as $a=a(1)\ldots a(L)$ where $a(\cdot)\in\{1,\ldots,n\}$ with corresponding affine contractions
\[ T_a(z)= p_a +\cf_a(z-p_a) \]
where it can be shown via induction \cite{ph00004} that $\cf_a\libeq \cf_{a(1)}\cdot\ldots\cdot\cf_{a(L)}$ with complex argument $\arg(\cf_a)\equiv \vt L\ (\mmod\ 2\pi)$ and $p_a\libeq T_a(0)/(1-\cf_a)$ where $T_a=T_{a(1)}\co\ldots\co T_{a(L)}$.

So at the iteration level $L_*\libeq M/\gcd(N,M)$ we have $\vt L_*\equiv 0\ (\mmod\ 2\pi)$ giving any map $T_a$ a zero rotation angle. Therefore $F$ is generated as a Sierpi\'{n}ski fractal by the $n^{L_*}$ IFS contractions $T_a$ of level $|a|=L_*$ implying that the convex hull is $\Conv(F)=\Conv(p_a: |a|=L_*)$. Thus if the extremal points $\Ext(p_a: |a|=L_*)$ lie on a circle, then the boundary $\partial\Conv(F)$ is a cyclic polygon, implying a circumcircle for the polyfractal $F$ generated by the IFS $\{T_a: |a|=L_*\}$.

Finding a circumcircle in the non-equiangular case, perhaps when $\Ext(p_1,\ldots,p_n)$ lie on a circle, is left open to the reader.

\section{Bounding Spheres}

\subsection{A General Bounding Sphere}

If we consider the circumcircle problem in a broader sense as the problem of bounding the attractor of any IFS in $\R^d\ (d\in\N)$, it hinges on the containment property
\[ \mH(B(c,r))\subset B(c,r)\ \ \mathrm{where}\ \ B(c,r)\libeq \{z\in\C:\ \|z-c\|_2\le r\} \]
which implies inductively for any level that $B(c,r)\supset\mH^L(B(c,r))\rightarrow F\ (L\tti)$.

Let us now take more general IFS contractions $T_k(z)=p_k+M_k(z-p_k)$ where $z,p_k\in\R^d$ and $M_k\in\R^{d\times d},\ \vl_k\libeq \|M_k\|_2<1$ in the matrix norm induced by the Euclidean norm. The map $\mH$ will still have an attractor $F$ \cite{ba00007}. The above containment property becomes
\[ \|T_k(c)-c\|_2 + \vl_k r \le r\ \ (k=1,\dots,n) \]
a weaker form of the tangential conditions for trifractals. Rewriting the left side and estimating it using these new notations, we get
\[ \vr(z)\libeq \max_{1\le k\le n} \|p_k-z\|_2\ \ (z\in\R^d)\ \ \mathrm{and}\ \ \vl_*\libeq \max_{1\le k\le n} \|M_k\|_2,\ \mu_*\libeq \max_{1\le k\le n} \|I-M_k\|_2 \]
\[ \|(I-M_k)(p_k-c)\|_2 + \vl_k r \le \|I-M_k\|_2\ \|p_k-c\|_2 + \vl_* r \le \mu_* \vr(c) + \vl_* r. \]
So to satisfy the containment property optimally, we require $\mu_* \vr(c) + \vl_* r = r$ implying that $r = r(c)\libeq \mu_*\vr(c)/(1-\vl_*)$. So taking any $c\in\R^d$, such as the centroid of $p_1,\ldots,p_n$, the closed ball $B(c,r(c))$ will be a bounding sphere of the fractal $F$.

Clearly a minimizer of $\vr(\cdot)$ also minimizes the corresponding radius $r(\cdot)$. This minimizer is known to be unique, so denote it as $c_*\libeq \mamin_{c\in\R^d}\ \vr(c)$. The corresponding sphere $B(c_*,\vr(c_*))$ is called the ``minimal bounding sphere'' of $p_1,\ldots,p_n\in\R^d$, and its determination is called the ``Smallest Bounding-Sphere Problem'', first investigated in modern times by Sylvester \cite{ba00016} in 1857. Several algorithms exist for finding the exact parameters of the optimal sphere, and the fastest run in linear time, such as the algorithms of Fischer et al. \cite{ic00008}, Larsson \cite{bc00007}, Megiddo \cite{ba00014}, and Welzl \cite{ba00015}.

If we wish to improve the tightness of a bounding sphere, it can be done easily by exploiting the self-similarity of $F=\mH(F)$. Having such a sphere $C=B(c,r)$ and taking its $L$-level iterate $\mH^L(C)$, we can compute the minimal bounding sphere $B(c',r')$ of the $n^L$ centers $\mH^L(\{c\})$, and then $B(c',r'+\vl_*^Lr)$ will be a tighter bounding sphere of $F$. For large enough $L$, we can get within any $\ve>0$ accuracy of the fractal.

Nevertheless, one might wonder how the circumcircle compares in the plane to the one derived above. According to our numerical experiments for bifractals, in about $2/3$ of randomized cases the circumcircle has a smaller radius, but this general bounding circle still remains competent, and in a few cases it is even tighter than the circumcircle.
\begin{figure}[H]
\centering
\includegraphics[width=400pt]{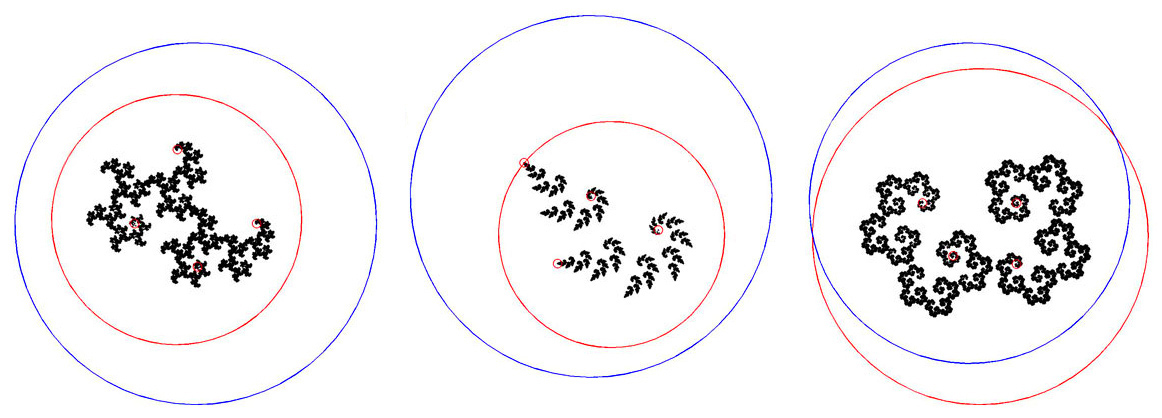}
\caption{The circumcircle (red) vs. the general bounding circle (blue) for two contractions.}
\end{figure}
Philosophically speaking, the above ``general bounding sphere'' reduces the problem of bounding an IFS fractal with an infinite number of points, to that of a finite number of points, the fixed points of the IFS.

\subsection{Other Bounds}

Dubuc and Hamzaoui \cite{br00003} find a bounding circle similar to our last one, but it remains unclear how the optimal center may be found. Rice \cite{bc00001} introduces a method for $n$-map IFS, similar to the circumcircle definition of this paper, but also relying on an optimization algorithm. Canright \cite{ba00002} gives an algorithmic method as well. Sharp et al. \cite{br00001, br00002} determine bounding circles with a given fixed center for the purpose of fitting the attractor on the screen. Martyn \cite{ba00004} gives an algorithm that seeks the tightest bounding sphere of an IFS fractal, via some potentially expensive subroutines.

As noted earlier, tightness can be improved to an arbitrary accuracy by further iteration, making this aspect of bounding less relevant. The circles and spheres introduced in this paper are special in that they are given by explicit formulas, unlike those in the literature.

\section{Concluding Remarks}

The reader may have found the formulation of equiangular polyfractals as a Sierpi\'{n}ski fractal somewhat peculiar, as it also implies the finiteness of extrema. The question arises if this can be shown in general; meaning is it true that any IFS fractal has a finite number of extremal points? This is answered by the author in the paper \cite{bu00010} focused on the determination of the convex hull of IFS fractals, which seems like a natural inquiry regarding bounding.

Indeed bounding $F$ by some invariant compact set $S\supset\mH(S)$ is a key prerequisite of various algorithms for IFS fractals -- such as the ray tracing of 3D IFS fractals \cite{bc00005} -- and the convex hull is often ideal in terms of efficiency \cite{ph00004}.

\bibliographystyle{abbrv}
\bibliography{mybib}
\addcontentsline{toc}{section}{\textbf{References}}

\end{document}